\documentclass[10pt,a4paper]{article}

\usepackage[T1]{fontenc}
\usepackage[utf8]{inputenc}
\usepackage{lmodern}
\usepackage{geometry}
\usepackage{graphicx}
\usepackage{booktabs}
\usepackage{tabularx}
\usepackage{array}
\usepackage{amsmath,amssymb}
\usepackage{caption}
\usepackage{hyperref}
\usepackage{enumitem}
\usepackage{float}

\graphicspath{{figures/}}
\setlist{nosep}
\hypersetup{colorlinks=true,linkcolor=black,citecolor=black,urlcolor=blue}
\title{\bfseries Cross-Sensor SAR Data Generation Using Diffusion Models\\and Feature Migration}
\author{WU Xuanting, ZHANG Fan, MA Fei\thanks{Corresponding author. E-mail address: \href{mailto:mafei@mail.buct.edu.cn}{mafei@mail.buct.edu.cn}.}\\
YIN Qiang, ZHOU Yongsheng\\
\small College of Information Science and Technology, Beijing University of Chemical Technology, Beijing 100029, P. R. China}
\date{}

\begin{document}
\maketitle

\begin{abstract}
Different synthetic aperture radar (SAR) sensors vary significantly in resolution, polarization modes, and frequency bands, making it difficult to directly apply existing models to newly launched SAR satellites. These new systems require large amounts of labeled data for model retraining, but collecting sufficient data in a short time is often infeasible. To address this contradiction, this paper proposes a data generation and transfer framework, integrating a stable diffusion model with attention distillation, that leverages historical SAR data to synthesize training data tailored to the unique characteristics of new SAR systems. Specifically, we fine-tune the low-rank adaptation (LoRA) modules within the multimodal diffusion transformer (MM-DiT) architecture to enable class-controllable SAR image generation guided by textual prompts. To ensure that the generated images reflect the statistical properties and imaging characteristics of the target SAR system, we further introduce an attention distillation mechanism that transfers sensor-specific features, such as spatial texture, speckle distribution, and structural patterns, from real target-domain data to the generative model. Extensive experiments on multi-class aircraft target datasets from two real spaceborne SAR systems demonstrate the effectiveness of the proposed approach in alleviating data scarcity and supporting cross-sensor remote sensing applications.
\end{abstract}

\noindent\textbf{Key words:} synthetic aperture radar (SAR); generative technology; cross-sensor; feature migration


\setcounter{section}{-1}
\section{Introduction}

Synthetic aperture radar (SAR), as an active microwave remote sensing technology, demonstrates significant advantages in the field of Earth observation, owing to its all-weather, all-day imaging capabilities and its sensitivity to the unique electromagnetic scattering characteristics of the Earth's surface. However, SAR systems also have inherent limitations. On one hand, the exceptionally large volume of SAR data leads to high costs for data processing and analysis. On the other hand, the utilization rate of existing data is relatively low, which makes the analysis and application of data across different platforms difficult \cite{shen2025lowcost}. This is primarily due to the differences in parameters among SAR sensors, such as imaging geometry, operating frequency, and polarization mode \cite{deng2024polsar}.

In the context of deep learning methods increasingly dominating SAR image interpretation tasks, generative data augmentation has become a key approach to alleviate the problem of limited labeled data. Researchers have proposed various SAR data generation and augmentation strategies, including generative adversarial networks (GANs) \cite{goodfellow2014gan}, diffusion models (DMs) \cite{ho2020ddpm}, and autoregressive models (ARs) \cite{vandenOord2016pixel}. Although these methods have enriched the technical approaches for SAR data augmentation to some extent, most are confined to expansion within single and homogeneous datasets \cite{tian2024weighted}. These approaches fail to effectively utilize the rich semantic information contained in other sensors or data sources, thereby limiting the generalization capability and adaptability of SAR data augmentation techniques to diverse scenarios \cite{kong2025fewshot}.

At the same time, with the continuous emergence of new-generation, high-resolution, and multi-mode spaceborne SAR systems, the problem of newly launched satellites being unable to accumulate sufficient labeled data to support effective deep learning model training in their early mission phases has become increasingly significant \cite{ma2022fastsar}. This directly leads to a lag in the research and development of intelligent interpretation algorithms for this new high-quality data, preventing the full realization of its application potential. How to rapidly generate high-quality, diverse training samples for new SAR systems has become a pressing challenge that needs to be addressed.

\begin{figure}[H]
  \centering
  \includegraphics[width=0.52\linewidth]{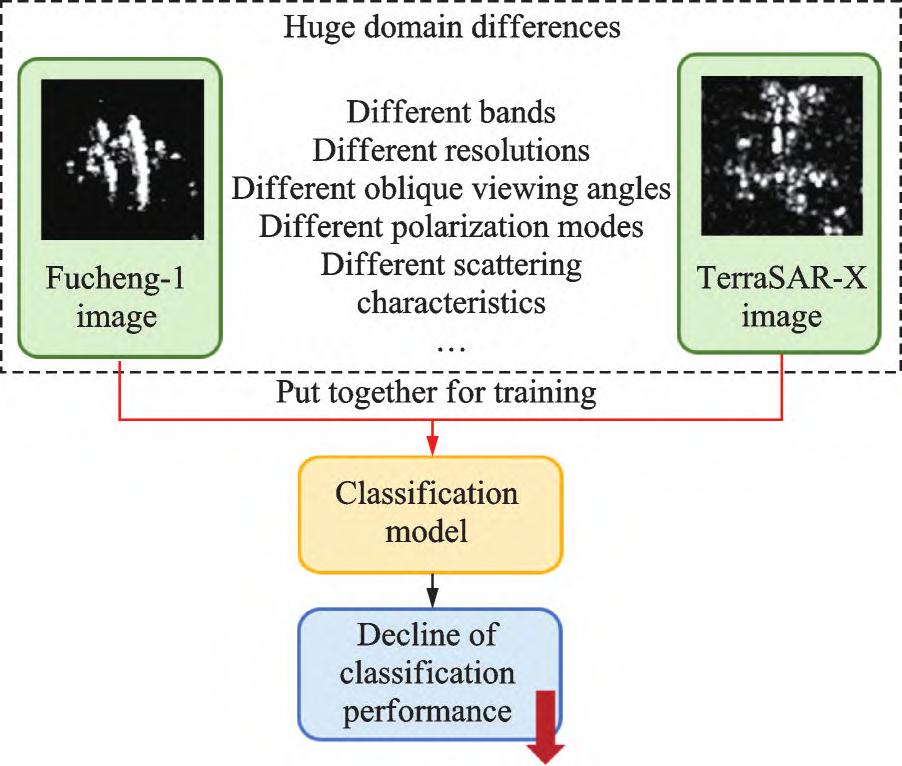}
  \caption{Classification accuracy degradation due to cross-sensor domain gaps.}
  \label{fig:domain-gap}
\end{figure}

For convenience we will refer to the old existing satellite payloads as SAR1, and the new satellite payloads launched will be referred to as SAR2. As illustrated in Fig.~\ref{fig:domain-gap}, images of the same target acquired by different SAR payloads exhibit significant visual discrepancies, posing a great challenge to the direct, cross-platform utilization of data. In Fig.~\ref{fig:domain-gap}, the red lines represent the training process and the red downward arrow represents a decline. To address the aforementioned challenges, the paper proposes a new framework that combines diffusion models and attention distillation techniques. This framework leverages the powerful image generation capabilities of diffusion models and the advantages of attention distillation in handling domain differences to achieve knowledge extrapolation from existing, abundant SAR1 data sources to SAR2 with scarce data, thereby generating high-quality SAR images that possess the imaging characteristics of SAR2.

\begin{figure}[H]
  \centering
  \includegraphics[width=0.52\linewidth]{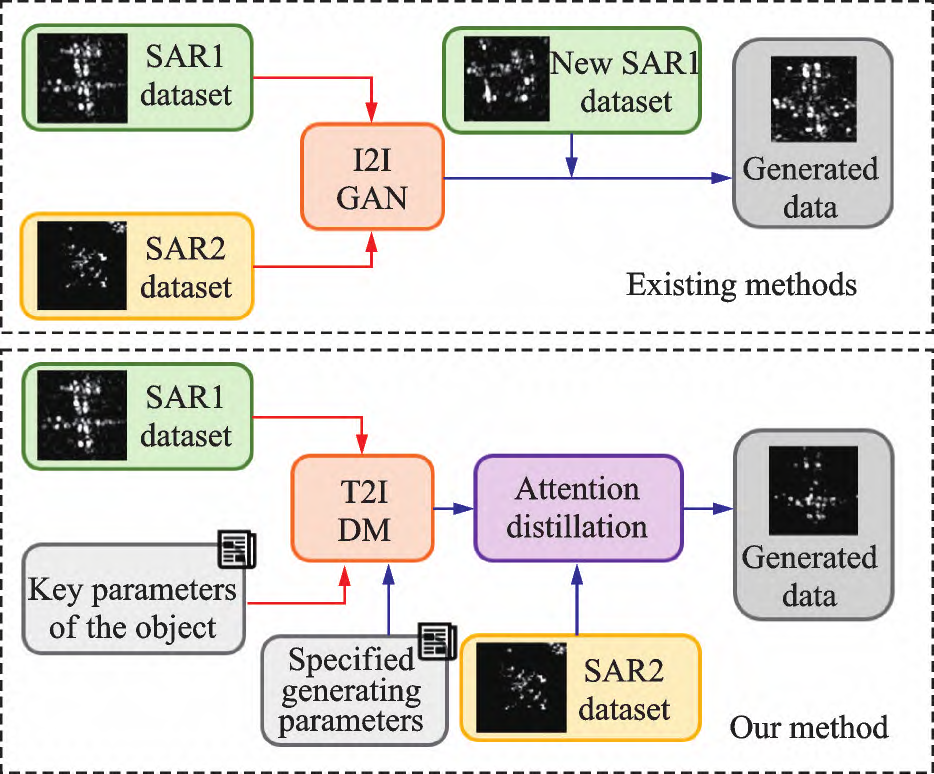}
  \caption{Differences between the proposed method and existing methods.}
  \label{fig:method-comparison}
\end{figure}

Fig.~\ref{fig:method-comparison} shows structural differences between the proposed approach and existing methods, in which the red and blue lines represent the training and inference processes, respectively. ``I2I GAN'' and ``T2I DM'' stand for image-to-image generative adversarial network and text-to-image diffusion model, respectively. Fig.~\ref{fig:method-comparison} highlights the key methodological shift from direct image-to-image (I2I) translation used in existing methods to the proposed two-stage framework, which decouples content generation from style transfer. While previous GAN methods rely on image-to-image generation, they can only generate SAR2 images from existing SAR1 image counterparts. Our method, by extracting SAR1 images, is able to generate an unlimited amount of content and requires only a small number of SAR2 images as feature guidance.

The core innovations of this paper are as follows.

\begin{enumerate}[label=(\arabic*)]
  \item We introduce a novel technical paradigm in the field of remote sensing image generation that fuses a large-scale text-to-image model with an attention distillation feature transfer method. This approach guides image generation using textual information and then, through attention feature transfer, injects the unique imaging style and physical statistical properties of the target sensor into the generated content. This unique combination, which decouples ``content'' and ``style'', effectively enables efficient, high-fidelity data transfer from a data-abundant satellite to a data-scarce satellite under unpaired conditions. This framework provides a viable method for enhancing the data diversity and availability for new SAR systems.
  \item To strengthen the learning of SAR's physical characteristics, we have modified the attention distillation mechanism. This is achieved by introducing a speckle statistical loss and a frequency domain loss to constrain the properties of the SAR images.
  \item To empirically validate the effectiveness of the proposed method, we construct a heterogeneous dataset containing aircraft targets from two real spaceborne SAR sensors. Experimental results demonstrate that the proposed framework possesses cross-domain generation capabilities.
\end{enumerate}

\section{Related Works}

\subsection{Optical-SAR conversion methods}

To address the issue of SAR data scarcity, an important research direction involves generating SAR imagery from the vast amount of available optical remote sensing imagery, a process known as optical-to-SAR conversion. This technical approach is primarily dominated by two categories of methods \cite{ienco2019combining}. Early research employed sophisticated convolutional neural network (CNN) architectures to establish a non-linear mapping between optical and SAR images at the feature level, such as using multi-branch networks or heterogeneous Siamese networks for feature fusion \cite{ye2020f3net}. To overcome the reliance of supervised learning methods on large numbers of registered image pairs, subsequent research shifted towards GANs, particularly image-to-image translation models like CycleGAN \cite{zhu2017cyclegan} and Pix2pix \cite{isola2017pix2pix}. For instance, Fu et al. \cite{fu2022optical} designed a multi-stage cascaded GAN framework to improve the quality of optical-to-SAR conversion.

Despite providing a pathway for data augmentation, the optical-to-SAR conversion approach has fundamental bottlenecks. The imaging mechanisms of optical and SAR sensors are fundamentally different, in which the former captures the spectral reflectance of ground objects, while the latter reflects geometric structure and dielectric properties. This significant modality gap makes it extremely difficult for models to generate SAR images with realistic physical characteristics. For example, physical phenomena crucial for SAR target recognition, such as speckle noise, layover, shadow, and specific scattering center intensities, are often distorted. Furthermore, the reliance of CNN methods on paired data \cite{schmitt2018sen12} and the inherent problems of GANs, such as training instability and susceptibility to mode collapse \cite{shen2025lowcost}, also limit practical value of this approach.

\subsection{SAR-SAR conversion methods}

To avoid the problem of physical distortion in cross-modality conversion, a more direct approach is to perform data extrapolation within the SAR modality itself, known as SAR-to-SAR conversion. This direction is also primarily driven by various generative models. GAN-based methods, such as CycleGAN, CUT \cite{park2020cut} and StarGAN \cite{choi2018stargan}, have become the mainstream choice for achieving style transfer between different SAR sensors due to their ability to handle unpaired data. For example, Pan et al. \cite{pan2020sar} successfully used CycleGAN to perform conversion between TerraSAR-X and Sentinel-1 data, enhancing the performance of downstream tasks. Meanwhile, autoregressive models, an emerging paradigm, draw on the success of large language models, generating images through ``next-token prediction''. Studies like SimpleAR \cite{wang2025simplear} have demonstrated their advantages in training stability and scalability, offering potential for high-fidelity, controllable generation.

However, these SAR-to-SAR methods also have their limitations. On one hand, the adversarial training process of GANs is unstable, and the cycle-consistency loss they rely on, being an indirect constraint, may distort or lose critical electromagnetic scattering features in the SAR images during the conversion process. On the other hand, the technology of AR models in the visual generation domain is not yet mature. Studies such as Selftok \cite{hong2024selftok} have pointed out that the visual tokenization paradigm on which current AR models depend has fundamental flaws and is not a true autoregressive structure, which limits the model's inference capabilities. At the same time, its token-by-token generation characteristic leads to unacceptable inference latency. While parallelization studies like PAR \cite{lee2024parallel} can accelerate the process, they do not fundamentally solve the problem.

In recent years, diffusion models have become the latest technical benchmark, surpassing GANs due to their excellent generation quality and stable training process, and have been successfully validated in the field of SAR image generation. The research by Qosja et al. clearly indicates that in both qualitative and quantitative comparisons, a standard DDPM significantly outperforms various GAN methods in SAR image synthesis tasks \cite{qosja2024sar}. To further enhance realism, researchers have explored different approaches. For example, Qosja et al. \cite{qosja2024sar} also verified that pre-training on large-scale unlabeled clutter data can effectively improve the quality of subsequent target generation. DiffuSAR \cite{ying2024diffusar}, in contrast, approaches the problem from physical characteristics of SAR images, emphasizing the importance of high-frequency (HF) components and significantly improving the quality of generated images by introducing a frequency conditioning module.

Although diffusion models have achieved great success in single-domain SAR image generation, there remain clear gaps and challenges in applying them to the more complex tasks of cross-sensor data extrapolation. Existing research has focused on generating high-fidelity samples within single, homogeneous datasets, and their model designs lack an effective mechanism specifically built for cross-domain style transfer that can decouple content and style. Furthermore, there is currently no mature solution for how to use the complex, high-dimensional imaging features of another SAR sensor as a precise condition to guide the generation process of a diffusion model. For instance, while DiffDet4SAR \cite{zhou2024diffdet4sar} uses the diffusion paradigm, its objective is object detection rather than style transfer. These challenges are core problems that this research work aims to solve.

\section{Methods}

To address the challenges of data scarcity for new SAR sensors and imaging differences across sensors, this paper proposes a generative framework based on ``content-style'' decoupling. This paradigm conceptually separates the physical structure of a scene (content) from the unique imaging signature of a specific sensor, which includes its speckle statistics and resolution-dependent textures (style). This decoupled design allows each stage to focus on a single, well-defined objective, permitting the adoption of the most advanced and suitable techniques for each sub-task: controllable content generation and high-fidelity feature transfer. Consequently, this approach achieves precision in feature migration under an unpaired setting while ensuring content controllability, thereby effectively solving the challenge of cross-sensor data extrapolation.

Let $D_s=\{x_i,c_i\}$ represent the large, annotated dataset from SAR1, where $x_i$ is a SAR image and
\[
c_i=\{\mathrm{SAR\ Image}, \mathrm{Class}\}
\]
is its corresponding textual description. Let $D_t=\{x_{\mathrm{sty}}\}$ be the scarce, unlabeled dataset from SAR2, in which a single reference image $x_{\mathrm{sty}}$ serves as the ground truth for the target imaging characteristics we aim to replicate.

\subsection{Overall framework}

\begin{figure}[H]
  \centering
  \includegraphics[width=\linewidth]{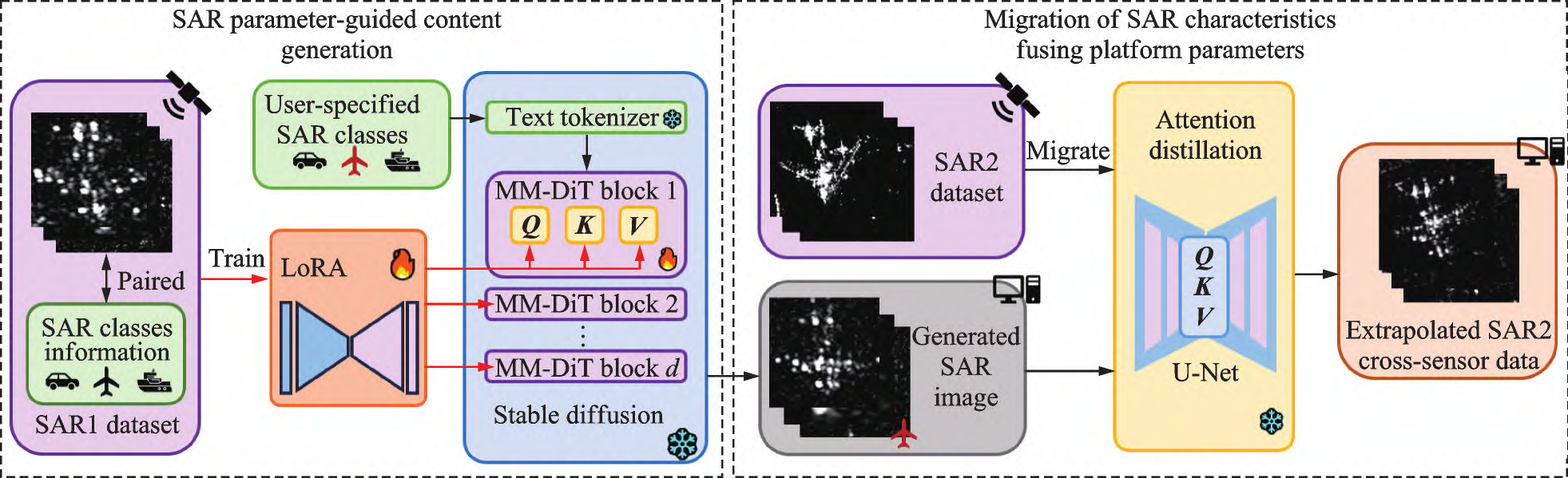}
  \caption{Framework for cross-sensor SAR data generation.}
  \label{fig:framework}
\end{figure}

This framework operates in two distinct stages. It first utilizes a text-to-image (T2I) diffusion model to generate target content with precise semantic and structural control based on textual prompts. Subsequently, through a novel optimization process founded on attention distillation, it precisely transfers the unique imaging characteristics of the target sensor, such as its speckle statistics, textural patterns, and radiometric response, onto this content base, ensuring high-fidelity style emulation in an unpaired setting. Its overall workflow is illustrated in Fig.~\ref{fig:framework}, in which the red and gray lines represent the training and inference processes, respectively.

SAR parameter-guided content generation utilizes a large amount of data from SAR1, along with its corresponding text information and key imaging parameters, to perform parameter-efficient fine-tuning on a stable diffusion 3.5 model \cite{chen2024scaling} using LoRA \cite{hu2022lora}. Once trained, this model is capable of generating content-controllable SAR images based on new text prompts.

The migration of SAR characteristics fusing platform parameters takes the ``content base'' generated in the first stage as a structural reference and requires only a small amount of data from SAR2 as the style target. Through a feature transfer method based on attention distillation \cite{cao2023masactrl}, the unique imaging characteristics of SAR2 are precisely injected into the content base, ultimately completing the cross-domain SAR image extrapolation.

\subsection{SAR parameter-guided content generation}

\begin{figure}[H]
  \centering
  \includegraphics[width=0.4\linewidth]{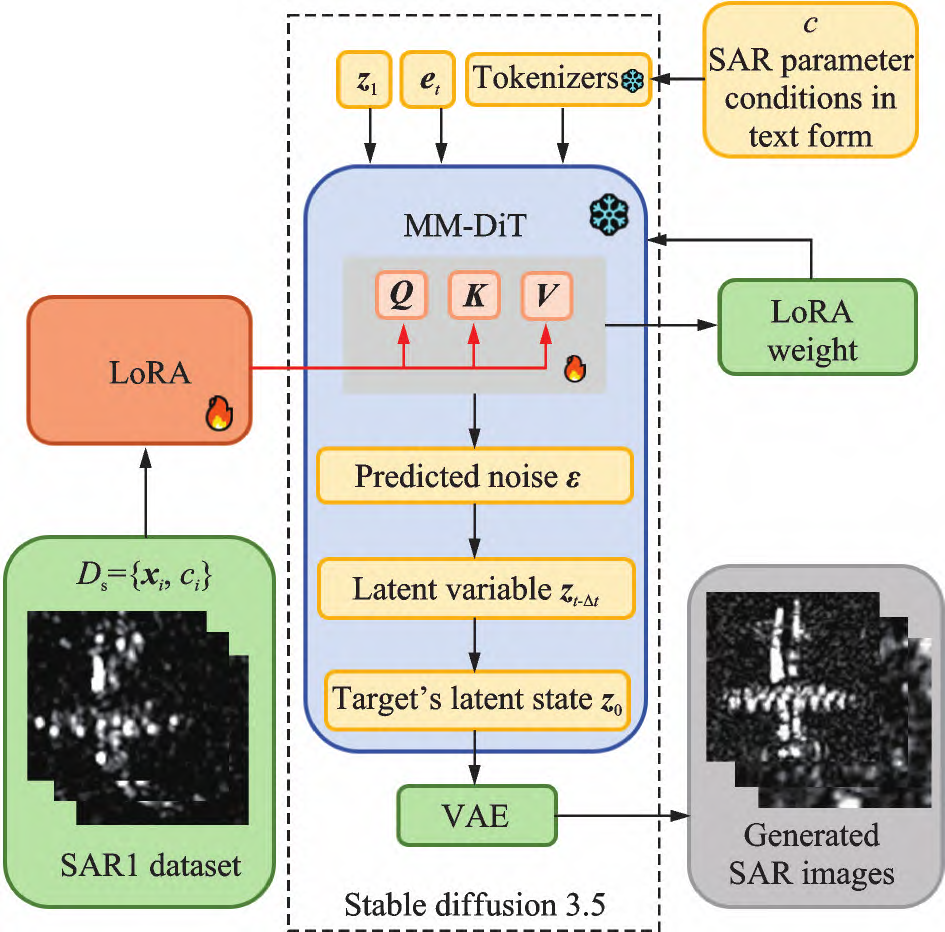}
  \caption{SAR parameter-guided content generation part of the proposed approach.}
  \label{fig:content-generation}
\end{figure}

The objective of this stage is to efficiently and controllably generate a ``content base'' that conforms to the physical structure of SAR scenes by utilizing text prompts. The workflow of the generation part of this framework is illustrated in Fig.~\ref{fig:content-generation}, in which the red and gray lines represent the training and inference processes, respectively. LoRA-fine-tuned multimodal diffusion transformer (MM-DiT) enables diffusion-based synthesis constrained by SAR parameters and text conditions.

The core of this stage is the adoption of an advanced T2I generative model, which serves as the foundation for subsequent LoRA fine-tuning. We choose to model our approach on the technology used in stable diffusion 3.5. The overall generation process is designed to progressively transform a random noise vector into a high-fidelity image under the guidance of a text prompt.

The entire process begins by sampling a pure Gaussian noise vector $z_1$ in the latent space. Subsequently, the model employs an iterative reverse denoising process, using a numerical ordinary differential equation (ODE) solver to advance step-by-step from timestep $t=1$ towards $t=0$. At each timestep $t$, the MM-DiT network takes the current noisy latent variable $z_t$, the embedding of the timestep $e_t$, and a pre-encoded text condition $c$ as input. Its task is to predict the noise contained within $z_t$. This predicted noise $\epsilon$ is then used to compute the next, less noisy latent variable $z_{t-\Delta t}$. After a preset number of iterative steps, the process ultimately yields a latent representation $z_0$ that contains all the information of the target image. Finally, $z_0$ is fed into the decoder of a pre-trained variational autoencoder (VAE), which reconstructs it from the latent space into the final full-pixel image. This model, based on the rectified flow (RF) \cite{liu2023flow} and MM-DiT architecture, demonstrates strong performance in both generation quality and text comprehension.

Unlike traditional denoising diffusion probabilistic models (DDPMs), RF is an emerging generative model paradigm that directly constructs a mapping from a noise distribution to a data distribution via an ODE. It connects noise from a standard normal distribution, $\epsilon \sim \mathcal{N}(0,I)$, and real data $x_0$ with a straight-line path in the latent space. Any point along this path can be parameterized as
\begin{equation}
z_t=(1-t)x_0+t\epsilon .
\end{equation}
where $t$ is the timestep. This straight-line trajectory, compared to the curved paths of traditional diffusion models, theoretically possesses superior properties, enabling a more efficient and stable sampling process.

The model's training objective is to learn a velocity field $v_\Theta$, but in practice, this process is typically reparameterized into a network $\epsilon_\Theta$ that predicts the noise $\epsilon$. Its simplified training objective can be expressed as a weighted mean squared error loss function
\begin{equation}
\mathcal{L}
=\mathbb{E}_{t,x_0,c,\epsilon}
\left[
w(t)\left\|\epsilon_\Theta(z_t,t,c)-\epsilon\right\|_2^2
\right],
\end{equation}
where $c$ is the conditional information such as text, and $w(t)$ is a loss weight related to the timestep. By optimizing this objective, the model $\epsilon_\Theta$ learns to predict the original noise $\epsilon$ from any given noise-corrupted sample $z_t$.

\begin{figure}[H]
  \centering
  \includegraphics[width=0.32\linewidth]{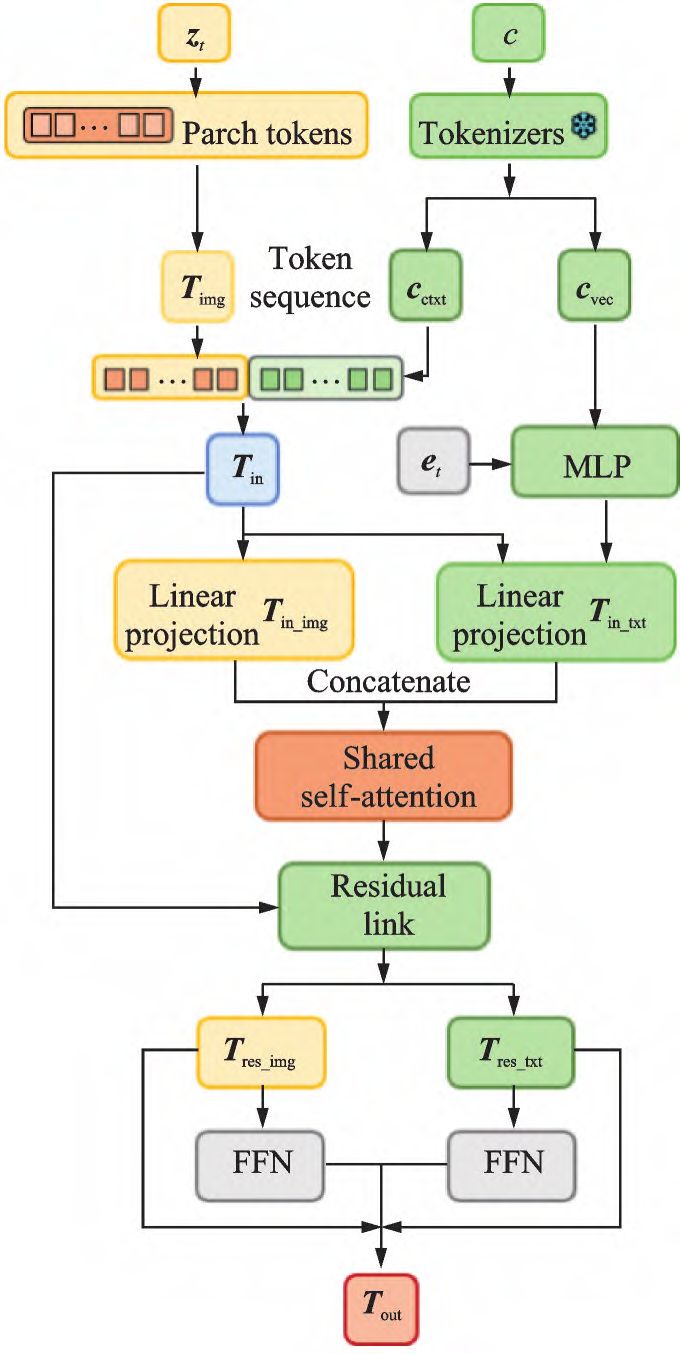}
  \caption{The core working principle of MM-DiT from accepting external SAR parameters in the form of text and random noise to inferring an image.}
  \label{fig:mmdit}
\end{figure}

MM-DiT is the core architecture of the model $\epsilon_\Theta$. Its core innovation is a dual-stream design that uses two separate sets of weights to process image and text tokens independently. Within each block of the transformer, these two token sequences are concatenated to share a single attention mechanism, which enables a bidirectional flow of information between the modalities. This architecture significantly improves the model's text comprehension, typography, and spatial reasoning, leading to higher-quality image generation that surpasses previous transformer-based backbones. The core workflow of MM-DiT is illustrated in Fig.~\ref{fig:mmdit}.

The input SAR image is first passed through the encoder of a VAE and compressed into a low-dimensional latent representation $x_0$. Subsequently, following the rectified flow formula described above, noise is added to obtain $z_t$. To be compatible with the Transformer architecture, $z_t$ is divided into $N_{\mathrm{img}}$ patch tokens and linearly projected into a sequence $T_{\mathrm{img}}$. The input text prompt $c$ is processed by multiple pre-trained and frozen text encoders to extract its semantic information, which is then differentiated into two forms of embeddings: a pooled vector for global guidance, $c_{\mathrm{vec}}$, and a token sequence containing detailed context, $c_{\mathrm{ctxt}}$. The timestep $t$ is converted into an embedding vector, $e_t$, through methods such as sinusoidal positional encoding.

MM-DiT discards the cross-attention mechanism found in traditional U-Nets, instead designing separate sets of Transformer weights for the image and text modalities. The input data is processed by $d$ stacked MM-DiT blocks. Within each processing block of MM-DiT, the image token sequence $T_{\mathrm{img}}$ and the text token sequence $c_{\mathrm{ctxt}}$ are concatenated along the sequence dimension to form a unified input sequence $T_{\mathrm{in}}$. Concurrently, the global text vector $c_{\mathrm{vec}}$ and the timestep embedding $e_t$ are fed into a small multi-layer perceptron (MLP) to generate the scaling parameters $\gamma$ and shifting parameters $\beta$ for the adaptive layer normalization (AdaLN):
\begin{align}
T_{\mathrm{in}} &= \mathrm{Concat}(T_{\mathrm{img}},c_{\mathrm{ctxt}}),\\
(\gamma,\beta) &= \mathrm{MLP}\bigl(\mathrm{Concat}(c_{\mathrm{vec}},e_t)\bigr).
\end{align}

Before the concatenated token sequence $T_{\mathrm{in}}$ is fed into the self-attention layer, its image and text portions are first processed by two separate linear projection layers. The two sets of processed tokens are then re-concatenated and input into a shared self-attention layer, shown as
\begin{align}
T'_{\mathrm{img}},T'_{\mathrm{txt}}
&= \mathrm{Linear}_{\mathrm{img}}(T_{\mathrm{in\_img}}),
   \mathrm{Linear}_{\mathrm{txt}}(T_{\mathrm{in\_txt}}),\\
\mathrm{Attn}_{\mathrm{out}}
&= \mathrm{SelfAttn}\left(
\mathrm{Linear}_{\mathrm{qkv}}\left(
\mathrm{Concat}(T'_{\mathrm{img}},T'_{\mathrm{txt}})
\right)\right).
\end{align}
This design allows information to interact bidirectionally between the feature streams of the image and text modalities, greatly enhancing the model's comprehension of complex text prompts.

The output of the attention layer $\mathrm{Attn}_{\mathrm{out}}$ is added to the input $T_{\mathrm{in}}$ through a residual connection. The image and text portions are then separated and processed separately by two independent feed-forward networks (FFNs), shown as
\begin{align}
T_{\mathrm{res}} &= \mathrm{Attn}_{\mathrm{out}}+T_{\mathrm{in}},\\
T_{\mathrm{img\_ffn}},T_{\mathrm{txt\_ffn}}
&=\mathrm{FFN}_{\mathrm{img}}(T_{\mathrm{res\_img}}),
  \mathrm{FFN}_{\mathrm{txt}}(T_{\mathrm{res\_txt}}).
\end{align}
Finally, they are again added to their respective inputs, completing the process of one MM-DiT block:
\begin{equation}
T_{\mathrm{out}}=
\left(
T_{\mathrm{img\_ffn}}+T_{\mathrm{res\_img}},
T_{\mathrm{txt\_ffn}}+T_{\mathrm{res\_txt}}
\right).
\end{equation}

After being deeply processed by $d$ MM-DiT blocks, the model only extracts the token portion corresponding to the image, $T_{\mathrm{out\_img}}$. This part of the sequence, after passing through an unpatching and linear projection layer, yields the final predicted noise, shown as
\begin{equation}
\epsilon=\mathrm{Linear}_{\mathrm{final}}\bigl(\mathrm{Unpatchify}(T_{\mathrm{out\_i}})\bigr),
\end{equation}
where the predicted noise $\epsilon$ is the single output of the entire $\epsilon_\Theta$ model. During inference, $\epsilon$ is used to guide the reverse denoising process to progressively generate the final image.

LoRA is a parameter-efficient fine-tuning technique that adapts large models to new tasks with minimal computational overhead. It achieves this by injecting trainable low-rank matrices into specific layers of a pre-trained model to approximate the effect of full parameter fine-tuning.

In the implementation of LoRA, for a pre-trained model's original weight matrix $W_0$, its state is kept frozen throughout the fine-tuning process and does not participate in gradient updates. The weight update, $\Delta W$, is approximated by the product of two much smaller low-rank matrices, $A$ and $B$, where the rank $r$ is far lower than the primitive dimension $d$ and $k$. Therefore, the weight update can be represented as
\begin{equation}
\Delta W=BA .
\end{equation}
During the training, only matrices $A$ and $B$ are trainable parameters. In the forward pass, for a given input vector $x$, the final output $h$ of a LoRA-adapted layer is the sum of the output from the original pre-trained path and the output from the low-rank adaptation path, which is shown as
\begin{equation}
h=W_0x+BAx .
\end{equation}

This design enables effective adjustment of the model's behavior without altering the original model's architecture or parameter count. In our framework, we apply LoRA adapters to the core of the SD3.5 model's MM-DiT architecture: the self-attention and cross-attention modules. A standard attention layer maps an input token sequence $T$ to Query, Key, and Value using three separate linear projection matrices. Our fine-tuning strategy is to freeze the original weight and introduce separate, trainable low-rank adapters for each. Therefore, after LoRA fine-tuning, the effective weight of each projection matrix in the attention module is composed of the original weight $W_i$ plus its corresponding low-rank update $B_iA_i$, which can be uniformly expressed as
\begin{equation}
W_i'=W_i+B_iA_i,\qquad i\in\{Q,K,V\}.
\end{equation}
By training only these low-rank matrices, which represent a minimal fraction of the total model parameters, we can precisely adjust how the model attends to and combines image and text features during the generation process. This targeted adaptation strategy concentrates limited training resources specifically on the self-attention and cross-attention blocks within the MM-DiT, which are the components most influential for semantic and structural modeling. This method thereby achieves highly efficient model fine-tuning, drastically reducing training time and memory requirements while yielding a small, portable adapter for the base model that can be easily deployed.

\subsection{Migration of SAR characteristics fusing platform parameters}

\begin{figure}[H]
  \centering
  \includegraphics[width=0.49\linewidth]{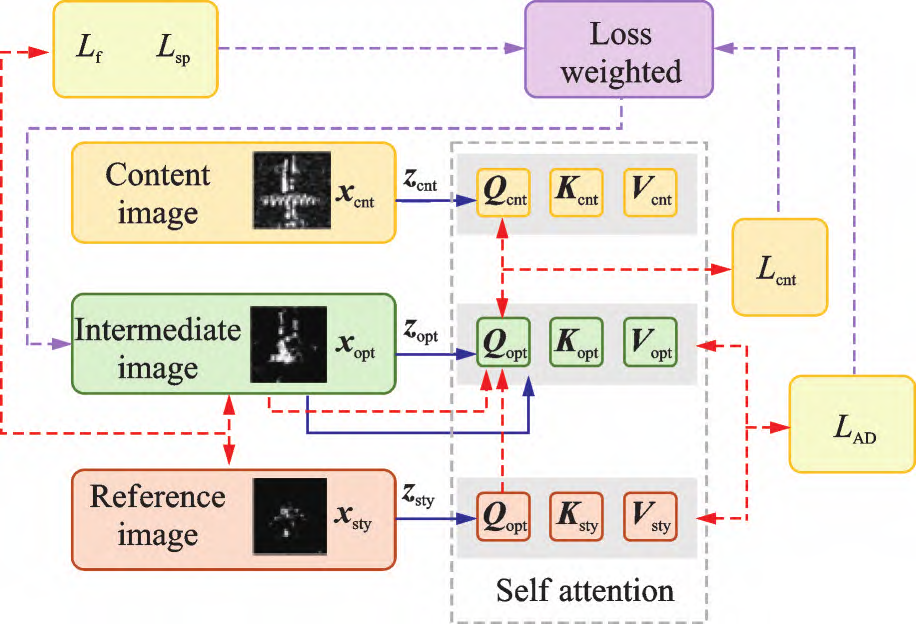}
  \caption{Migration of SAR characteristics fusing platform parameters module.}
  \label{fig:attention-distillation}
\end{figure}

The second stage of this framework, migration of SAR characteristics fusing platform parameters, transfers the unique imaging characteristics from a single, unpaired SAR image of the target sensor onto the content base. The attention distillation method used in this framework is illustrated in Fig.~\ref{fig:attention-distillation}, in which the red, blue, and purple lines represent the training process, the inference process of the latent space, and the loss calculation process, respectively.

The self-attention layers of the pre-trained U-Net in a diffusion model capture rich, multi-scale visual features that collectively define an image's ``style''. In the attention distillation process, a real SAR image from the target sensor is first encoded as $z_{\mathrm{sty}}$:
\begin{equation}
z_{\mathrm{sty}}=E(I).
\end{equation}
In each optimization step, we simultaneously feed the image code to be optimized, $z_{\mathrm{opt}}$, and the style reference code, $z_{\mathrm{sty}}$, into the pre-trained U-Net and extract features from its self-attention layers. In specific layers of the network, we compute two attention outputs: ideal style output $A_{\mathrm{ideal}}$ and current style output $A_{\mathrm{current}}$.

$A_{\mathrm{ideal}}$ is computed using $Q_{\mathrm{opt}}$ from $z_{\mathrm{opt}}$ and the $K_{\mathrm{sty}}$ and $V_{\mathrm{sty}}$ from $z_{\mathrm{sty}}$. If $z_{\mathrm{opt}}$ possesses the features of $z_{\mathrm{sty}}$, its attention output would be
\begin{equation}
A_{\mathrm{ideal}}=\mathrm{SelfAttn}(Q_{\mathrm{opt}},K_{\mathrm{sty}},V_{\mathrm{sty}}).
\end{equation}
$A_{\mathrm{current}}$ is computed using $Q_{\mathrm{opt}}$, $K_{\mathrm{opt}}$, and $V_{\mathrm{opt}}$ from $z_{\mathrm{opt}}$ itself, representing the current stylistic state of $z_{\mathrm{opt}}$, shown as
\begin{equation}
A_{\mathrm{current}}=\mathrm{SelfAttn}(Q_{\mathrm{opt}},K_{\mathrm{opt}},V_{\mathrm{opt}}).
\end{equation}

We design four loss functions to better transfer the characteristics of SAR2 onto the content map of SAR1. The AD loss is used to calculate the style gap between the style map and the content map; the content loss is used to maintain the content features of the original SAR1 image; and the speckle statistical loss and frequency loss are used from the perspective of physical characteristics to assist in fine-tuning the physical feature migration of the SAR image.

The AD loss $L_{\mathrm{AD}}$ is the core of attention distillation feature transfer, distilling visual elements by re-aggregating features within the self-attention mechanism. The AD loss is defined as the sum of the $L_1$ distance between these two outputs across all selected layers, shown as
\begin{equation}
L_{\mathrm{AD}}=\left\|A_{\mathrm{ideal}}-A_{\mathrm{current}}\right\|_1.
\end{equation}
We define content loss $L_{\mathrm{cnt}}$ similarly to AD loss, also based on the self-attention mechanism, fully leveraging the diffusion model's advantage of deep understanding of images. Specifically, by calculating the $L_1$ loss between the target query $Q_{\mathrm{opt}}$ and the reference query $Q_{\mathrm{cnt}}$, we obtain the content loss as
\begin{equation}
L_{\mathrm{cnt}}=\left\|Q_{\mathrm{opt}}-Q_{\mathrm{cnt}}\right\|_1.
\end{equation}

To replicate SAR-specific characteristics, we develop a targeted strategy for selecting the U-Net's self-attention layers for distillation. We simultaneously select layers from both the deeper and shallower stages of the U-Net. Deeper layers capture macroscopic structure and semantic layout, ensuring the target's overall contour remains stable. Shallower, higher-resolution layers are sensitive to local textures, edges, and noise, enabling the replication of the target sensor's unique speckle distribution, textural details, and scattering sidelobes. By performing distillation at these different depths, our method synergistically transfers both the macro-structure and micro-texture of the target SAR sensor, generating highly realistic extrapolated data.

While attention distillation can effectively capture and replicate the multi-scale visual features of a target sensor's images, this matching occurs primarily in the deep feature space and does not explicitly guarantee that the generated images adhere to the underlying physical and statistical laws of SAR imaging. To address this issue, we introduce two constraint terms based on SAR physical priors.

Speckle noise is a core statistical characteristic of SAR images \cite{zhong2025scattering}. Therefore, to ensure statistical fidelity, we introduce a speckle statistical loss that aligns the speckle intensity of the generated and reference images by matching their equivalent number of looks (ENL) \cite{ma2022region}. ENL is a classic metric for measuring speckle intensity in homogeneous regions of an image, which is defined as
\begin{equation}
\mathrm{ENL}=\frac{\mu^2}{\sigma^2},
\end{equation}
where $\mu$ represents the image mean and $\sigma$ the image standard deviation. In each optimization step, we randomly select $M$ patches of size $N\times N$ from the homogeneous background regions of the reference image $x_{\mathrm{sty}}$. We select $M$ corresponding patches from the image being optimized, $x_{\mathrm{opt}}$. We then compute the average ENL values for both sets of patches and define the speckle statistical loss as the $L_1$ distance between them:
\begin{equation}
L_{\mathrm{sp}}=
\left\|\overline{\mathrm{ENL}}(x_{\mathrm{opt}})
-\overline{\mathrm{ENL}}(x_{\mathrm{sty}})\right\|_1.
\end{equation}
By incorporating $L_{\mathrm{sp}}$ into the total loss function, we directly guide the optimization process to ensure that the speckle fluctuations in the generated image are statistically consistent with the target sensor's real data.

Studies like DiffuSAR have shown that the HF components of SAR images are crucial for replicating their scattering details. Therefore, our method introduces a frequency domain loss to align the energy distribution in the high-frequency components of the generated and reference images. In each optimization step, we apply a 2D fast Fourier Transform to both the current generated image $x_{\mathrm{opt}}$ and the reference image $x_{\mathrm{sty}}$ to obtain their frequency spectra. We then define an annular high-pass filter mask, $M_{\mathrm{HF}}$, which has a value of 0 at the center and 1 in the surrounding regions. The frequency domain loss is defined as the $L_1$ distance between their high-frequency components, shown as
\begin{equation}
L_f=\left\|
\mathcal{F}(x_{\mathrm{opt}})\odot M_{\mathrm{HF}}
-\mathcal{F}(x_{\mathrm{sty}})\odot M_{\mathrm{HF}}
\right\|_1.
\end{equation}
This loss term directly forces the spectral structure of the generated image, particularly the high-frequency details critical for SAR target recognition, to match that of the target sensor's real data. By incorporating these physical priors, the final optimization objective for feature extrapolation is
\begin{equation}
L_{\mathrm{total}}
=L_{\mathrm{AD}}+\lambda_{\mathrm{cnt}}L_{\mathrm{cnt}}
+\lambda_{\mathrm{sp}}L_{\mathrm{sp}}+\lambda_fL_f .
\end{equation}
where $\lambda_{\mathrm{cnt}}$ represents content loss weight, $\lambda_{\mathrm{sp}}$ speckle statistical loss weight, and $\lambda_f$ frequency loss weight. In our experiments, the loss weights are set based on empirical analysis to balance the three objectives of content retention, style transfer, and physical realism. The weight $\lambda_{\mathrm{cnt}}$ of the content loss $L_{\mathrm{cnt}}$ is set to 0.25 to ensure the structural fidelity of the generated images. The weights $\lambda_{\mathrm{sp}}$ and $\lambda_f$ for the speckle statistical loss $L_{\mathrm{sp}}$ and frequency domain loss $L_f$ are both set to 0.1. These two physical prior terms serve as a supplement and fine-tuning to the attention distillation style loss $L_{\mathrm{AD}}$, guiding the model to generate details more consistent with SAR imaging principles without dominating the entire optimization process. These settings are proven effective in our experiments.

\section{Experiments}

\subsection{Experimental dataset and parameter introduction}

To comprehensively evaluate the performance of the proposed framework, we construct a heterogeneous spaceborne SAR dataset containing seven different types of aircraft targets. The heterogeneous dataset used in experiments is composed of data from two distinct SAR payloads, representing different generations of spaceborne technology. The source domain (SAR1) data is from Germany's established TerraSAR-X satellite, launched in June 2007, while the target domain (SAR2) data is from the more recent Fucheng-1 (Fuxi-1) satellite, launched in January 2023. These sensors operate under fundamentally different physical parameters, most notably in their frequency band and spatial resolution. TerraSAR-X operates in the high-frequency X-band and provides high-resolution imagery of approximately 1 m, making it highly sensitive to surface textures and fine details. In contrast, Fucheng-1 operates in the lower-frequency C-band with a moderate resolution of around 3 m, resulting in different penetration capabilities and less detailed textural information. This significant disparity in both operating frequency and resolution creates a substantial domain gap, leading to distinct backscattering characteristics, speckle statistics, and textural representations in the imagery from the two payloads. To mitigate the long-tail effect, the data from Fucheng-1 is controlled. The specific data quantities for the dataset are shown in Table~\ref{tab:dataset}.

\begin{table}[H]
\centering
\caption{Dataset details.}
\label{tab:dataset}
\small
\resizebox{\linewidth}{!}{%
\begin{tabular}{lrrrrrrr}
\toprule
Dataset name & B-52 & Boeing-747 & Boeing-767 & C-130 & C-17 & F-15 & F-16\\
\midrule
Fucheng-1 dataset & 50 & 50 & 50 & 50 & 50 & 50 & 50\\
TerraSAR-X dataset & 129 & 146 & 132 & 134 & 165 & 143 & 134\\
Dataset after adding generated data & 366 & 383 & 369 & 371 & 402 & 382 & 373\\
\bottomrule
\end{tabular}}
\end{table}

In the experiment, all comparative methods are used to learn the image style transfer from the TerraSAR-X to Fucheng-1, with the generated images serving as augmented data for Fucheng-1. All GAN-based comparative methods use similar training epochs and optimizer settings. All GAN models are trained for a total of 200 epochs. The batch size is uniformly set to 12. We select the Adam optimizer for both the generator and discriminator of all models, with an initial learning rate of $1\times10^{-4}$.

\subsection{Comparative methods}

To evaluate our framework, we select several representative GAN-based models from the I2I translation field for comparison. These include classic and modern unpaired methods such as CycleGAN, which uses cycle-consistency loss; StarGAN, which extends translation to multiple domains with a single generator; and the more recent contrastive unpaired translation (CUT), which employs contrastive learning for higher efficiency. Additionally, we select Pix2pix, a foundational model for paired I2I translation, as a supervised reference. These baselines are compared against our method, the complete two-stage framework based on content generation and feature extrapolation.

\subsection{Result of generation and migration}

In order to visually evaluate the performance of the proposed two-stage framework in cross-payload data extrapolation tasks, we qualitatively visualize the generation and migration results for different aircraft targets, as shown in Fig.~\ref{fig:generation-results}. The results are presented in the following figure. TerraSAR-X (Row 1) vs. Fucheng-1 (Row 2) imaging differences are resolved through generated content (Row 3) and final outputs (Row 4) that preserve TerraSAR-X aircraft structures while replicating Fucheng-1's speckle, brightness, and textures. The figure clearly demonstrates the complete transformation process from the source domain to the target domain and validates the capability of our approach in both content retention and style migration.

\begin{figure}[H]
  \centering
  \includegraphics[width=0.98\linewidth]{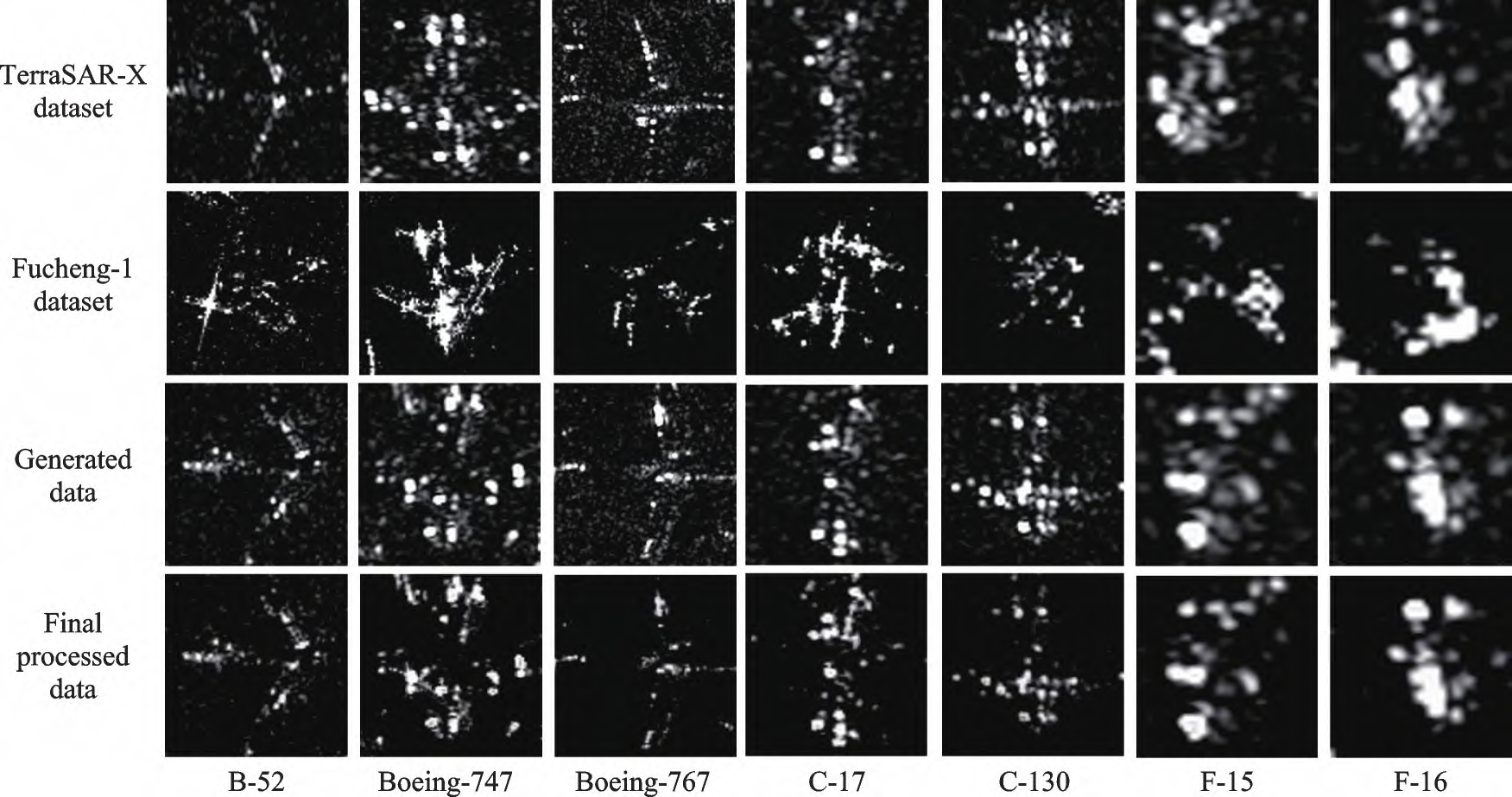}
  \caption{Real and synthetic SAR samples.}
  \label{fig:generation-results}
\end{figure}

\subsection{After adding the generated data}

\begin{figure}[H]
  \centering
  \includegraphics[width=0.52\linewidth]{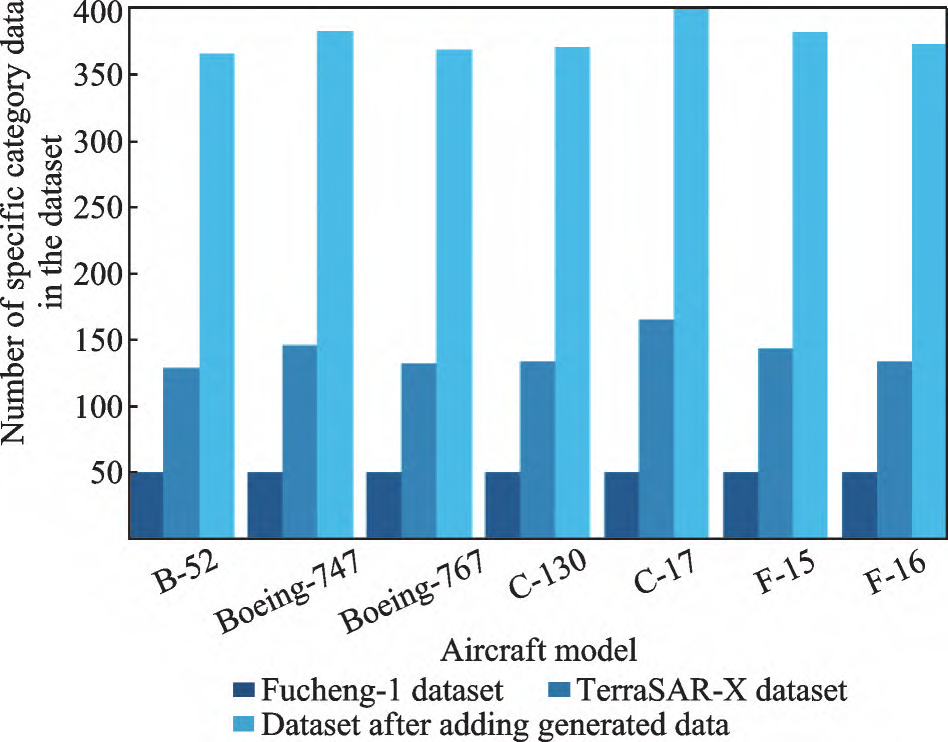}
  \caption{Histogram of distribution of each type of data on the dataset.}
  \label{fig:dataset-distribution}
\end{figure}

We use performance on a downstream classification task as the core evaluation metric for data augmentation effectiveness. Fig.~\ref{fig:dataset-distribution} shows the distribution of data quantities for each aircraft type in the dataset, demonstrating that our method can expand the quantity of effective data, thereby improving the performance of the classification network. We use a standard ResNet50 \cite{he2016resnet} classification network, which is trained on a mixed training set composed of ``a small amount of real data from the new sensor'' and ``augmented data generated by each method''. The model is then evaluated on a fixed test set of real data from Fucheng-1.

We utilize four metrics commonly used in classification performance studies: Accuracy, precision, recall and F1-score to evaluate the performance of each class of methods on the ResNet50 classification network.

\begin{figure}[H]
  \centering
  \includegraphics[width=0.48\linewidth]{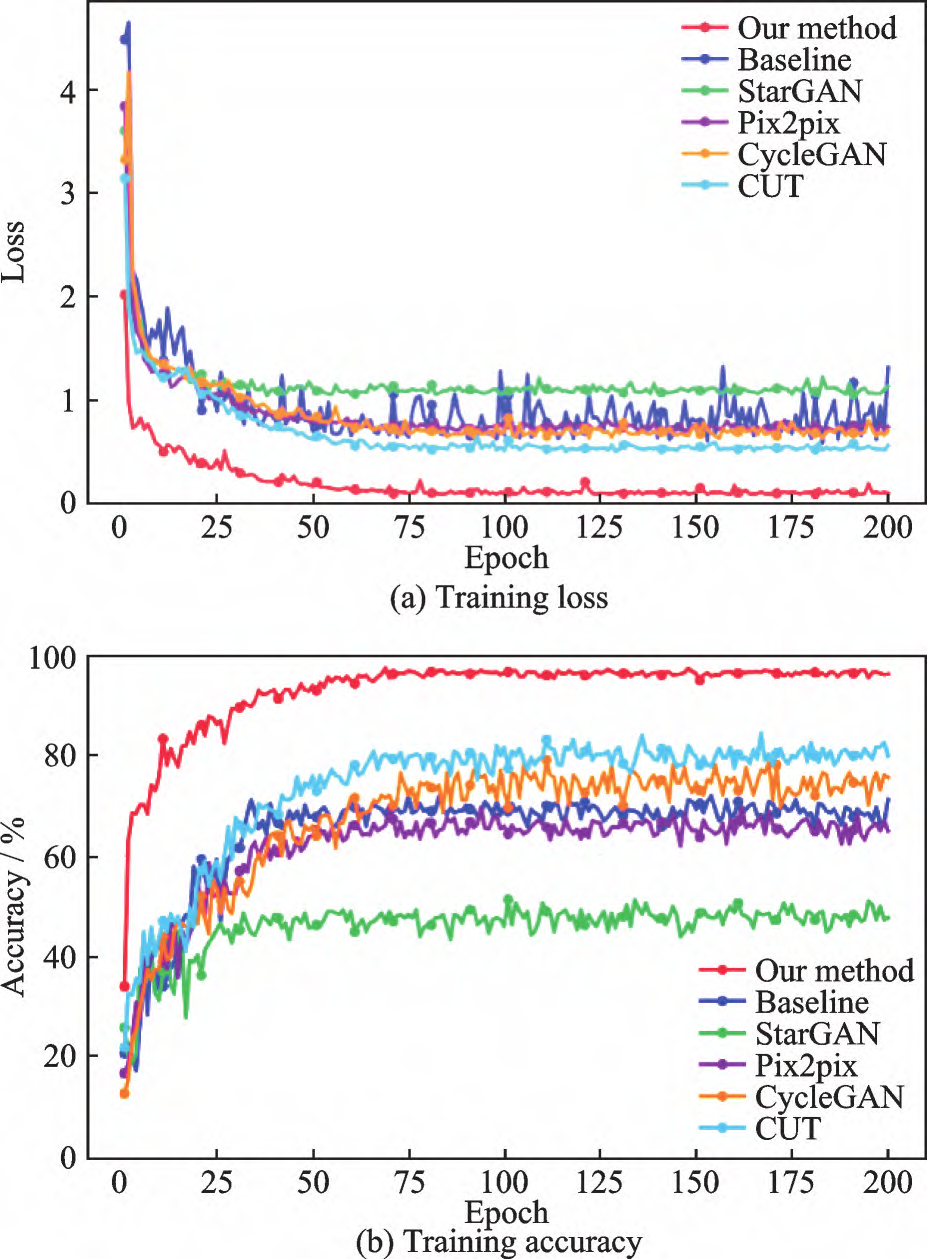}
  \caption{Training performance comparison.}
  \label{fig:training-curves}
\end{figure}

During the training phase of a deep learning model, training loss and training accuracy are two fundamental metrics used to monitor the learning process. Training loss is the primary value that the optimization algorithm directly minimizes and it is a numerical representation of the error, or discrepancy, between the model's predictions and actual ground-truth labels on the training dataset. This value, which is often calculated using functions like Cross-Entropy for classification, provides essential gradient information needed to update the model's weights. In parallel, training accuracy offers a more intuitive measure of performance, representing the percentage of training samples that the model has classified correctly. While training loss provides a fine-grained signal for optimization, training accuracy gives a high-level, human-interpretable summary of how well the model is fitting the data it has seen. Generally, as the training loss decreases, the training accuracy is expected to increase, indicating a successful learning progression. Fig.~\ref{fig:training-curves} represents these two key parameters profiles for each type of model when trained on ResNet50. It can be seen from Fig.~\ref{fig:training-curves} that our framework achieves superior convergence speed, stability, and final accuracy over GAN-based methods in both loss reduction and classification accuracy.

\begin{figure}[H]
  \centering
  \includegraphics[width=0.42\linewidth]{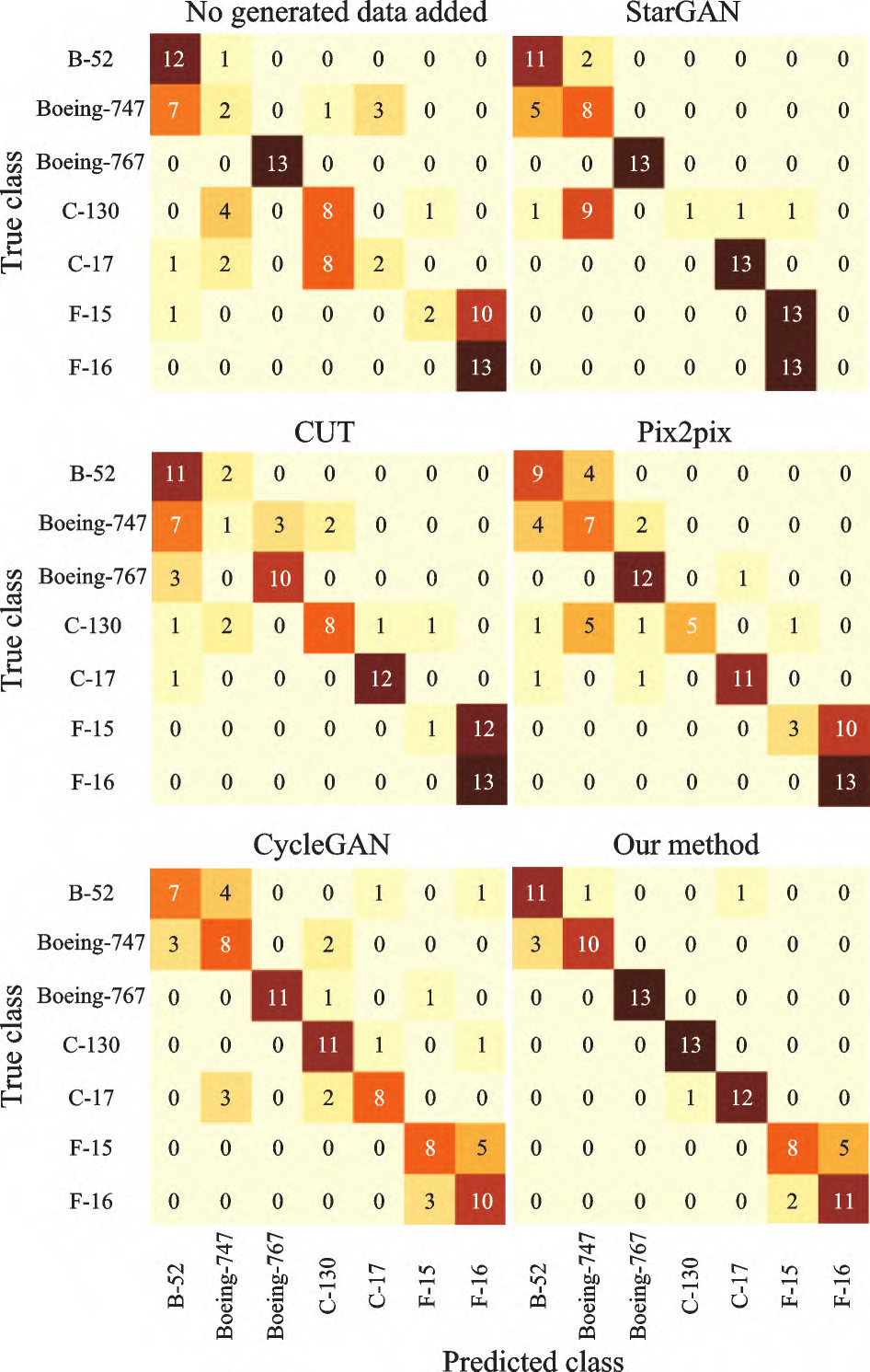}
  \caption{Confusion matrices for each comparative experimental group.}
  \label{fig:comparison-analysis}
\end{figure}

A confusion matrix is a specific table used to evaluate a classification model's performance; it lays out the counts of correct and incorrect predictions for each class, providing a detailed breakdown of exactly what kind of errors the model is making. Fig.~\ref{fig:comparison-analysis} shows a visual analysis of results of comparison experiments, using confusion matrices to visualize the performance of the downstream tasks.

The data generated by each method is used to enhance the training of the downstream classification model, with the performance results shown in Table~\ref{tab:comparison}. For ease of comparison, we also include the classification results of a baseline model, which is trained using only the small amount of real data from Fucheng-1.

\begin{table}[H]
\centering
\caption{Improvement of classification accuracy by adding data generated by different methods in the Fucheng-1 training set.}
\label{tab:comparison}
\small
\begin{tabular}{lrrrr}
\toprule
Method & Accuracy/\% & Precision/\% & Recall/\% & F1-score/\%\\
\midrule
No generated data added & 57.142 & 55.659 & 57.143 & 51.650\\
StarGAN & 50.549 & 37.165 & 50.549 & 41.328\\
CUT & 61.538 & 59.865 & 61.538 & 56.110\\
Pix2pix & 65.934 & 71.705 & 65.934 & 63.770\\
CycleGAN & 69.231 & 71.082 & 69.231 & 69.396\\
Our method & 85.714 & 86.199 & 85.714 & 85.549\\
\bottomrule
\end{tabular}
\end{table}

Analyzing experimental results, a detailed comparison reveals the varied efficacy of different data augmentation strategies when compared to the baseline model (57.14\% accuracy), which is trained only on the limited Fucheng-1 data. While some GAN-based image translation techniques show potential, their performance is inconsistent; notably, CycleGAN improves the accuracy to 69.23\%, demonstrating the viability of the unpaired translation concept. However, other adversarial methods struggle, performing even below the baseline, which underscores the significant challenge posed by the domain gap and the inherent instability of adversarial training for this specific task. In stark contrast, the proposed framework significantly surpasses all comparative methods across all evaluated metrics, achieving a final accuracy of 85.71\%. This represents a substantial improvement of over 16 percentage points compared to the best-performing method, CycleGAN. This superior performance is attributed to our framework's two-stage decoupled design, which generates a richer diversity of content via its T2I module and ensures higher fidelity style transfer through a more stable, optimization-based attention distillation process. This result conclusively demonstrates the capability and robustness of the proposed framework for cross-payload SAR data extrapolation.

\subsection{Ablation experiment}

To validate the effectiveness of each component of our framework, we conduct a series of ablation experiments. We continue to use the performance on a downstream classification task as the core evaluation metric for data augmentation effectiveness. To simulate the real-world scenario of data scarcity for a new sensor, the test set consistently uses 30\% of Fucheng-1 data, while the training set is composed according to different ablation settings. In research on small-sample learning and data-scarce scenarios, a 70\%/30\% training/test split is a common and widely accepted setting. It ensures sufficient data for model evaluation while maximizing the simulation of the challenge of insufficient training data. Our choice follows the general practice in this field to ensure the comparability of experimental results. Reducing the test set proportion (e.g., to 10\%) would lower the confidence of evaluation results; increasing it (e.g., to 50\%) would further exacerbate the scarcity of training data, potentially rendering all models (including baseline models) unable to be effectively trained, thereby making it difficult to distinguish the true performance differences introduced by different data augmentation methods. Therefore, 30\% is considered a balanced choice.

\begin{figure}[H]
  \centering
  \includegraphics[width=0.56\linewidth]{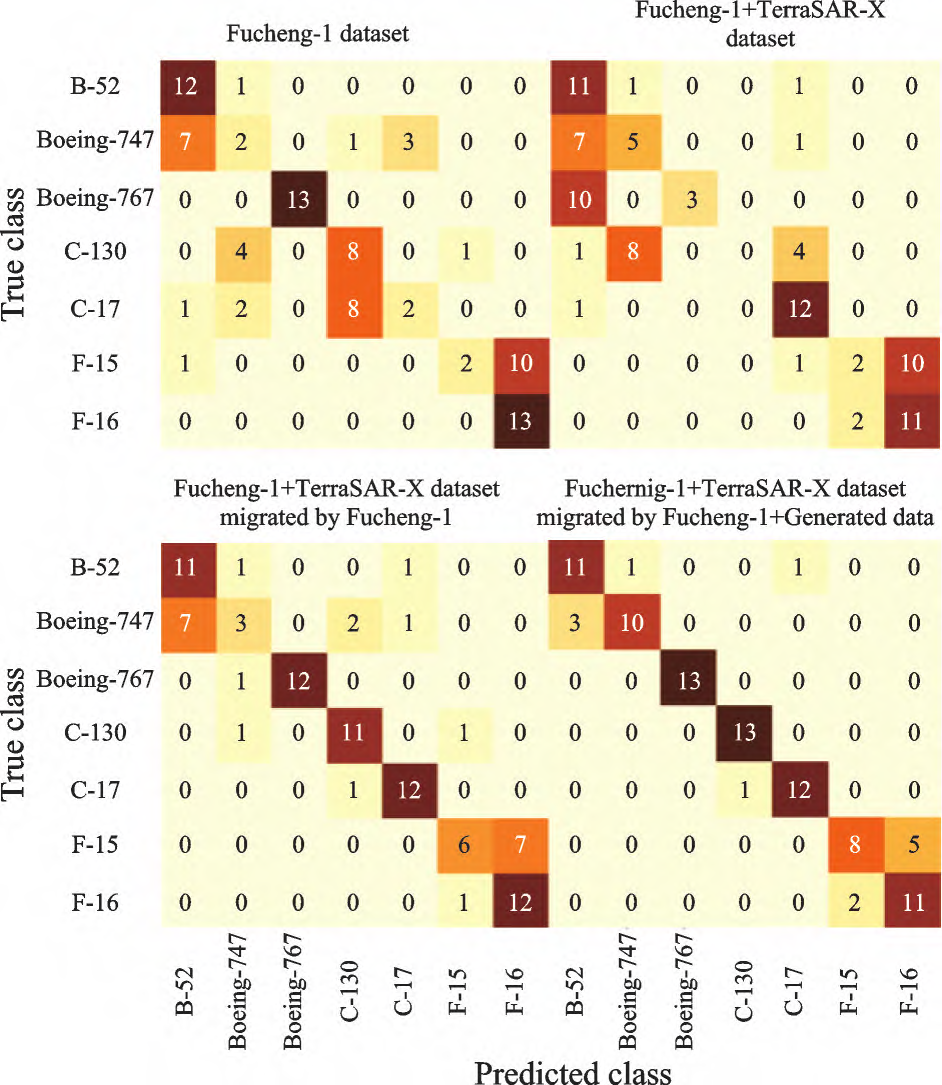}
  \caption{Confusion matrices for each ablation experimental group.}
  \label{fig:ablation-analysis}
\end{figure}

To verify the necessity of the proposed two-stage data extrapolation framework, we design four comparative experimental groups. Experimental results are shown in Table~\ref{tab:ablation}. Fig.~\ref{fig:ablation-analysis} shows a visual analysis of results of ablation experiments.

\begin{table}[H]
\centering
\caption{Ablation experiment results.}
\label{tab:ablation}
\small
\begin{tabularx}{\linewidth}{Xrrrr}
\toprule
Training dataset & Accuracy/\% & Precision/\% & Recall/\% & F1-score/\%\\
\midrule
Fucheng-1 dataset & 57.142 & 55.659 & 57.143 & 51.650\\
Fucheng-1+TerraSAR-X dataset & 48.351 & 48.275 & 48.352 & 41.276\\
Fucheng-1+TerraSAR-X dataset migrated by Fucheng-1 & 73.626 & 73.365 & 73.626 & 71.580\\
Fucheng-1+TerraSAR-X dataset migrated by Fucheng-1+Generated data & 85.714 & 86.199 & 85.714 & 85.549\\
\bottomrule
\end{tabularx}
\end{table}

The ablation study demonstrates the necessity of each component in our two-stage framework. The baseline model, trained only on the relative scarce Fucheng-1 dataset, achieves 57.14\% accuracy. Naively adding the TerraSAR-X data degrades performance to 48.35\%, confirming that a significant domain gap makes simple data pooling detrimental. By using only the style transfer component, which aligns the TerraSAR-X data with the Fucheng-1 style, the accuracy dramatically increases to 73.63\%. This result validates that our style extrapolation module is crucial for effectively bridging the domain gap. Finally, our complete framework, which combines diverse content generation with style transfer, achieves the superior accuracy of 85.71\%. This step-by-step improvement validates that both content generation and feature migration are indispensable, and their synergy is key to the framework's success.

To evaluate the specific contribution of the speckle statistical loss and the frequency domain loss, we conduct the following ablation analysis. This analysis aims to isolate and demonstrate the key role of these two physics-based prior constraints in enhancing the quality of cross-sensor SAR image generation. The results are shown in Table~\ref{tab:losses}.

\begin{table}[H]
\centering
\caption{Effectiveness analysis of losses.}
\label{tab:losses}
\small
\begin{tabular}{lrrrr}
\toprule
Index & Accuracy/\% & Precision/\% & Recall/\% & F1-score/\%\\
\midrule
Only with $L_{\mathrm{sp}}$ & 83.284 & 83.371 & 84.167 & 81.728\\
Only with $L_f$ & 82.893 & 81.644 & 83.543 & 80.469\\
With both losses & 85.714 & 86.199 & 85.714 & 85.549\\
\bottomrule
\end{tabular}
\end{table}

When the speckle statistical loss and the frequency domain loss work in conjunction, the framework's performance is optimal. This demonstrates a strong complementary relationship between them. The downstream classification model trained with this data generated by both losses exhibits the best performance, validating that these two loss functions are indispensable components for successfully achieving high-fidelity cross-sensor data extrapolation.

\section{Conclusions}

Addressing critical challenges of labeled data scarcity and significant imaging differences across sensors in new-generation spaceborne SAR systems, this paper has proposed and validated a novel cross-sensor SAR data extrapolation framework founded on ``content-style'' decoupling. The method first utilizes a large, pre-trained text-to-image model, which is fine-tuned with parameter-efficient LoRA on source domain data, to generate a semantically controllable content base representing the physical structure of the target. Subsequently, it employs an optimization-based attention distillation technique, which uniquely incorporates SAR-specific physical prior constraints such as speckle statistics and frequency domain characteristics, to precisely transfer the imaging style of the target sensor onto this content base under fully unpaired conditions. Experiments on a real-world, heterogeneous aircraft dataset demonstrate that the augmented data generated by our framework improves the accuracy of a downstream classification task to 85.71\%, significantly outperforming prominent GAN-based methods. This research not only provides an effective technical solution for the data bottleneck problem in new SAR systems, but also establishes a new paradigm for the cross-modal and cross-sensor intelligent interpretation of remote sensing images.

\section*{Acknowledgements}

This work was supported in part by the National Natural Science Foundations of China (Nos. 62201027, 62271034).

\section*{Authors}

The first author Mr. WU Xuanting received the B.S. degree in automation from Beijing University of Chemical Technology, Beijing, China, in 2024. He is currently working toward the master's degree in control science and engineering at the same university. His research interests include radar signal processing and deep learning generative tasks.

The corresponding author Dr. MA Fei received the B.S., M.S., and Ph.D. degrees in electronic and information engineering from Beijing University of Aeronautics and Astronautics (BUAA), Beijing, China, in 2013, 2016, and 2020, respectively. He is currently an associate professor at College of Information Science and Technology, Beijing University of Chemical Technology, Beijing. His research interests include radar signal processing, image processing, machine learning, and target detection.

\section*{Author Contributions}

Mr. WU Xuanting designed the study, compiled the models, conducted the analysis, interpreted the results, and wrote the manuscript. Prof. ZHANG Fan and Dr. MA Fei supervised the project and contributed to the overall study design and manuscript revision. Prof. ZHOU Yongsheng and Prof. YIN Qiang contributed to the discussion, provided key model components, and assisted with the data analysis. All authors commented on the manuscript draft and approved the submission.

\section*{Competing Interests}

The authors declare no competing interests.

\end{document}